\documentclass{WileyMSP-template}
\usepackage{amsmath}
\usepackage{graphicx}
\usepackage{epsfig}
\usepackage{latexsym,amssymb,bm}
\usepackage{epstopdf}
\usepackage{color}
\begin{document}

\pagestyle{fancy}
\rhead{\vspace{0.5cm}}

\title{Substrate-induced maximum optical chirality\\ 
of planar dielectric structures}


\maketitle


\author{Maxim V. Gorkunov}*
\author{Alexander A. Antonov}
\author{Alena V. Mamonova}
\author{Egor A. Muljarov}
\author{Yuri Kivshar}*


\dedication{}

\begin{affiliations}
Dr. Maxim V. Gorkunov\\
Shubnikov Institute of Crystallography, NRC ``Kurchatov Institute'', Moscow 119333, Russia\\
National Research Nuclear University MEPhI (Moscow Engineering Physics Institute), Moscow 115409, Russia; 
Email Address: gorkunov@crys.ras.ru\\

Dr. Alexander A. Antonov\\
Shubnikov Institute of Crystallography, NRC ``Kurchatov Institute'',Moscow 119333, Russia\\
Chair in Hybrid Nanosystems, Ludwig-Maximilians-University of Munich, Munich 80539, Germany\\

Ms. Alena V. Mamonova\\
Shubnikov Institute of Crystallography, NRC ``Kurchatov Institute'', Moscow 119333, Russia\\

Dr. Egor A. Muljarov\\
School of Physics and Astronomy, Cardiff University, Cardiff CF24 3AA, United Kingdom\\

Prof. Yuri Kivshar\\
Nonlinear Physics Centre, Australian National University, Canberra ACT 2601, Australia\\
Email Address: yuri.kivshar@anu.edu.au\\

\end{affiliations}



\begin{abstract}
	Resonant dielectric planar structures can interact selectively with light of particular helicity thus providing an attractive platform for {\it chiral flat optics}. The absence of mirror-symmetry planes defines geometric chirality,  and it remains the main condition for achieving strong circular dichroism. For planar optical structures such as photonic-crystal membranes and metasurfaces, breaking of out-of-plane mirror symmetry is especially challenging, as it requires to fabricate meta-atoms with a tilt, variable height, or vertically shifted positions. Although transparent substrates formally break out-of-plane mirror symmetries, their optical effect is typically subtle being  rarely considered for enhancing optical chirality. Here we reveal that low-refractive-index substrates can induce up to maximum optical chirality in otherwise achiral metastructures so that the transparency to waves of one helicity is combined with resonant blocking of waves of the opposite helicity. This effect originates from engineering twisted photonic eigenstates of different parities. Our perturbation analysis developed in terms of the resonant-state expansion reveals how the eigenstate coupling induced by a substrate gives rise to a pair of chiral resonances of opposite handedness. Our general theory is confirmed by the specific examples of light transmission in the normal and oblique directions by a rotation-symmetric photonic-crystal membrane placed on different transparent substrates.
\end{abstract}


\section{Introduction}

From early days of the systematic studies of optical chirality  \cite{Herschel1822, Fresnel1824}, it was established its unambiguous relation with geometric chirality, when a chiral object is distinguishable from its mirror image and cannot be superposed onto it~\cite{Kelvin1894}. Nowadays, this relation determines, in particular, the persistent improvement of precise optical instruments for quantifying weak molecular optical chirality, as most of biologically active molecules are geometrically chiral~\cite{Polavarapu2018}. Different types and classes of micro- and nanoscale structures have been designed to amplify weak natural optical chirality and to transfer its fingerprints to more convenient ranges~\cite{AvalosOvando2022}. At the same time, clear practical prospects of chiral photochemistry~\cite{He2018} and chiral quantum optics~\cite{Lodahl2017} motivate further the development of the tools of chiral optics able to generate, detect, and transform chiral light most efficiently.    

\medskip

Chiral metamaterials and, especially, their two-dimensional analogs--chiral metasurfaces--can manifest remarkable artificial optical chirality which, being by many orders of magnitude stronger than natural optical chirality, gives rise to a much broader palette of chiral optical effects. Following the link between geometric and optical chirality, subwavelength arrays of helices and springs were fabricated by means of sophisticated techniques in order to achieve large values of circular dichroism (CD) and optical rotation (OR) within subwavelength scales~\cite{Gansel2009,Singh2013,Gibbs2013,Kaschke2014,Esposito2016}.
In parallel, it was demonstrated that nanostructures of substantially simpler asymmetric shapes can provide even higher values of CD and OR \cite{Plum2007,Decker2010,Dietrich2014,Gorkunov2014,Kondratov2016,Zhu2017,Gorkunov2018,Tanaka2020}. 
Later, it was noticed that one can tailor optical chirality by adjusting the inner structure of photonic eigenstates: isolating a state from waves of a certain helicity creates the so-called {\it chiral quasi-bound state in the continuum} (quasi-BIC)~\cite{Gorkunov2020,Overvig2021}. This strategy has delivered a number of remarkably simple metastructures built as arrays of rods, bars, and pillars of highly refracting transparent materials \cite{Gorkunov2021, Zhang2022, Chen2023, Kuehner2023} eventually exhibiting up to maximum optical chirality, remaining transparent for waves of one helicity and blocking the waves of the opposite helicity~\cite{Fernandez-Corbaton2016}.

\medskip

For metasurfaces and other planar optical structures, chiral symmetry can be broken naturally by lifting in-plane and out-plane mirror symmetries. The former can be achieved in very different ways by implementing appropriate asymmetric two-dimensional patterns. Breaking the out-of-plane symmetry is more challenging for the modern lithography-based nanotechnology which is perfected for cutting precisely shaped nanoscale holes and slits with vertical walls in the layers of a constant thickness. On their own, the fabricated structures retain a mirror symmetry plane bisecting them through the middle. One has to complicate the technology by introducing additional steps in order to create tilted meta-atoms  \cite{Zhang2022,Chen2023} or meta-atoms of different heights \cite{Kuehner2023}.

\medskip

From a simple geometrical point of view, the presence of a transparent substrate breaks the out-of-plane symmetry regardless of the fabricated in-plane symmetry of the metasurface. Except for several particular studies~\cite{Powell2010,Albooyeh2015,Tian2020}, this fact is rarely recognized as a critical factor for the metasurface design. Conventional substrates are made of low-index transparent materials, and their effect on the metasurface optics is typically rather subtle. For chiral metasurfaces, it is not easy to separate a weak substrate-driven contribution determined by geometric chirality from the effect produced by the in-plane asymmetry~\cite{Toftul2024}. The latter can also provide specific selectivity on the metasurface interaction with waves of different helicity leading, for example, to strong CD of polarization conversion~\cite{Semnani2020,Shi2022}.  

\medskip

Here, we suggest and demonstrate a general strategy for employing  the symmetry-breaking effect of a metasurface substrate for achieving large (and even nearly maximum) optical chirality. We focus on metasurfaces possessing $C_4$ rotational symmetry which forbids polarization conversions for transmitting normally incident waves. The chirality is directly manifested by the transmission CD and OR, and the substrate-driven effects can be studied freely of other polarization features inherent to metasurfaces of lower symmetries~\cite{Gorkunov2024}. 

\medskip

The paper is arranged as follows. In Sec.~\ref{sec:idea}, we formulate the main concept, originating from the ideas of Ref.~\cite{Powell2010}, that the presence of a transparent weakly refracting substrate can cause strong effects, provided that the substrate may induce a coupling of the metasurface eigenstates of different spatial parity.  It is expected that the coupling is stronger when the eigenfrequencies of these states are close to each other. In combination with intrinsic degeneracy of the eigenstates of rotation-symmetric metasurfaces, this requires to consider four eigenstates in total. 

\medskip

In Sec.~\ref{sec:RSE}, we perform the detailed analysis in terms of the resonant-state expansion (RSE) theory \cite{Muljarov2010,Muljarov2018} which is known to be especially useful for evaluating optical effects caused by various weak perturbations of metasurface structure or environment~\cite{Weiss2017,Both2022,Almousa2023,Almousa2024}. By analyzing intrinsic CD of hybridized eigenstates in Sec.~\ref{sec:chiral} we identify the eigenstate parameters essential for achieving strong and eventually maximum optical chirality. As a particular illustration of the general ideas, in Sec.~\ref{sec:comsol} we present our numerical results for a specific case of optical membrane metasurface when a photonic-crystal slab (PCS) is placed on different transparent substrates. This PCS is a flat layer of weakly absorbing dielectric material perforated with a $C_4$ symmetric array of vertical rectangular holes. We verify that the substrate is responsible for the appearance of pairs of well separated {\it chiral transmission resonances} observed exclusively for left circularly polarized (LCP) and right circularly polarized (RCP) waves. The effect appears to be remarkably stable against  weak perturbations of the direction of the light incidence. We draw the main conclusions and discuss potential applications in Sec.~\ref{sec:concl}.


\section {Underlying concept}\label{sec:idea}

\begin{figure}
	\centering\includegraphics[width=1.0\textwidth]{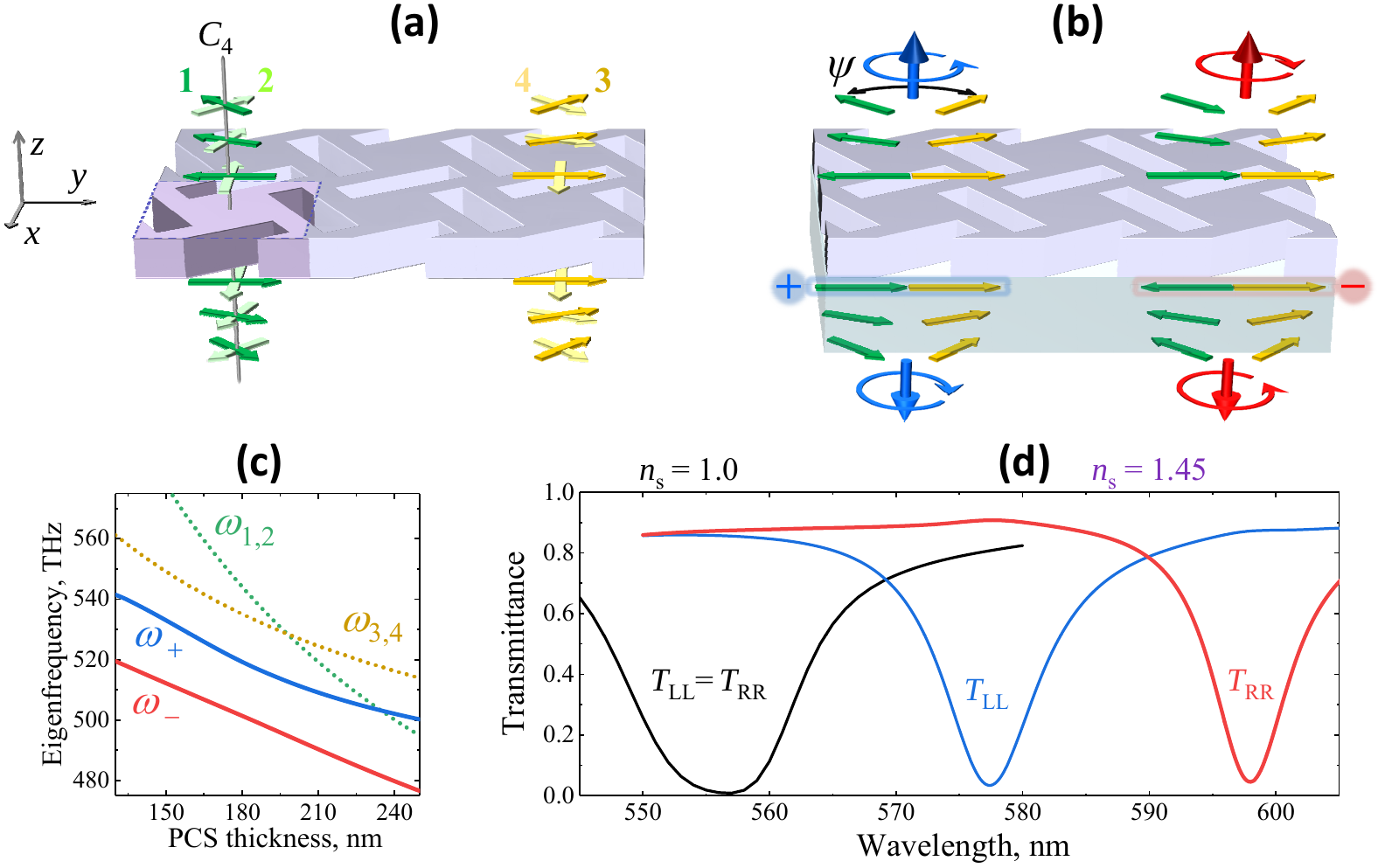} \caption{Main principle of maximum optical chirality arising as a result of eigenstate hybridization. (a) Pairs of degenerate eigenstates of a rotation-symmetric PCS in symmetric environment: odd sates 1 and 2 (on the left), and even states 3 and 4 (on the right). Coupling through the substrate transforms odd and even states into degenerate pairs of hybridized states. Examples of such states arising from merging of states 1 and 3 with  parallel (on the left) and antiparallel (on the right) orientation of the near electric field in the substrate are schematically shown in (b). Twisted structure of states 1 and 3 determines that the angle $\psi$ between the electric field of their far field asymptotics is close to $\pi/2$. (c) Real parts of eigenfrequencies of uncoupled ($\omega_{1,2}$ and $\omega_{3,4}$) and hybridized ($\omega_+$ and $\omega_-$) states numerically evaluated as functions of PCS thickness. (d) Simulated LCP and RCP transmission spectra of a PCS of a thickness of 180~nm suspended in a symmetric environment ($n_{\rm s}=1$) and upon a transparent glass-like substrate ($n_{\rm s}=1.45$). The PCS material refractive index is $n_{\rm PCS}=2.2+0.014 i$, see Section~\ref{sec:comsol} for all PCS geometric parameters.} \label{fig:1}
\end{figure}

As optical properties of dielectric metasurfaces are dominated by resonances underpinned by photonic eigenstates, we start with considering how the eigenstate symmetry is determined by the metasurface symmetry. First of all, we note that eigenstates of a $C_4$ rotation symmetric metasurface can be classified by their intrinsic rotational symmetry. Those, for which the coupling to free-space plane waves is allowed by the symmetry, always remain double degenerate \cite{Kondratov2016, Gorkunov2020, Sakoda1995, Hopkins2015, Doiron2022}. Indeed, applying the operator $\hat{R}$ of rotation by $\pi/2$ about the $z$-axis to an eigenstate (say, state 1) should produce the same state 1 or another eigenstate (state 2) having the same eigenfrequency. In the former case, state 1 appears to be incompatible with far-field plane waves, while in the latter case one can write  
\begin{eqnarray}
	{\hat{R}}{\bf F}_{1}({\bf r})={\bf F}_{2}(\hat{R}^{-1}{\bf r}),\label{rot1and2}\\
	{\hat{R}}{\bf F}_{2}({\bf r})=-{\bf F}_{1}(\hat{R}^{-1}{\bf r}),\label{rot2and1}
\end{eqnarray}
where we collect all eigenstate field distributions in a single super-vector ${\bf F}_n({\bf r})=\{{\bf E}_n({\bf r}),i{\bf H}_n({\bf r})\}$, and the minus sign in Equation~\eqref{rot2and1} ensures that applying $\hat{R}^2$, i.e., rotating by $\pi$, produces the same state but with the inverted sign of all field components. 

\medskip

Eigenstates of an achiral PCS in symmetric environment possess field distributions having specific parity with respect to the vertical $z$-axis. Although a transparent substrate is a weak perturbation, it can produce considerable observable consequences, if it induces otherwise absent coupling of eigenstates of different parity. To show this, we consider a pair of degenerate odd states 1 and 2 with $E_{1,2\ x,y}(x,y,z)=-E_{1,2\ x,y}(x,y,-z)$ and a pair of degenerate even states 3 and 4 with $E_{3,4\ x,y}(x,y,z)=E_{3,4\ x,y}(x,y,-z)$.  States 3 and 4 transform obeying the rotation rules similar to \eqref{rot1and2} and \eqref{rot2and1}. As different field components have different parity, we choose to classify the state parity by the parity of the in-plane electric field components which determine the coupling to normally incident plane waves. For more details of symmetry and parity properties of the states including  all field components, see SI Section~S1.

\medskip

In such terms, substrate-driven optical chirality can appear according to the scenario shown in Figure~\ref{fig:1}: as the substrate with a refractive index $n_{\rm s}>1$ breaks the PCS environment symmetry, states of different parity, shown in Figure~\ref{fig:1}(a), merge into parallel ($+$) and antiparallel ($-$) hybrids shown in Figure~\ref{fig:1}(b). We presume that, with a good accuracy, the far-field asymptotics of odd and even eigenstates of a PCS in symmetric environment is linearly polarized \cite{Zhen2014,Hsu2017}. 
Then, for the far-field asymptotics of hybrid states to be circularly polarized, the linearly polarized asymptotics of the unperturbed odd and even states should have the electric field amplitudes different by an angle $\psi\approx\pi/2$ in direction and by $\phi\approx\pi/2$ in phase, while having similar magnitudes.    

\medskip 
As we show below, it is possible to approach such conditions in a relatively simple PCS provided that the eigenfrequencies of its odd and even states are close enough. 
We use the PCS thickness as a natural controlling parameter: the frequencies of odd (1,2) and even (3,4) states vary differently with thickness and form a convenient crossing around the thickness of 200~nm, see Figure~\ref{fig:1}(c). In the presence of substrate, the coupling between the states transforms the crossing  into avoided crossing of two branches of parallel ($+$) and antiparallel ($-$) hybrid states. Remarkable results of such hybridization are shown in Figure~\ref{fig:1}(d): the achiral dip in the transmission underpinned by close odd and even states is transformed into a doublet of maximum-chiral transmission dips of opposite handedness.

\section{Substrate-induced mode hybridization}\label{sec:RSE}

The described origin of optical chirality is based on hybridization of a finite number of high-quality-factor metasurface eigenstates, which can be conveniently described in terms of the RSE. To apply the RSE to metasurfaces as open optical resonators, it is important to introduce the correct state normalization \cite{Muljarov2010,Muljarov2018}:
\begin{equation}
	1=||{\bf F}_n||^2=\int_V\left[\varepsilon({\bf r}){\bf E}_n\cdot{\bf E}_n-{\bf H}_n\cdot{\bf H}_n\right]dV+\frac{ic}{\omega_n}\int_{\partial V}\left[{\bf E}_n\times({\bf r}\cdot\nabla) {\bf H}_n +{\bf H}_n\times ({\bf r}\cdot\nabla){\bf E}_n\right] \cdot d{\bf S}, \label{norm}
\end{equation}
where the volume $V$ encloses the metasurface, $\partial V$ is its boundary surface, and all materials are supposed to be nonmagnetic with the permeability $\mu= 1$. Note that the field profiles of normalized states do not possess the usual dimensions of electric and magnetic fields.

\medskip

We use the first order RSE perturbation theory to express the hybridized states ${\bf \tilde{F}}_{n}$ as linear superposition of states ${\bf F}_{1-4}$ of the achiral PCS in symmetric environment. Important property of states ${\bf F}_{1-4}$ is their mutual orthogonality ensured by their symmetry. Indeed, one can introduce an analogue of scalar product defined as:
\begin{equation}
	({\bf F}_n\cdot{\bf F}_m)=\int_V[\varepsilon({\bf r}){\bf E}_n({\bf r})\cdot{\bf E}_m({\bf r}) -{\bf H}_n({\bf r})\cdot{\bf H}_m({\bf r})]dV+\frac{ic}{\omega_n-\omega_m}\int_{\partial V}\left[{\bf E}_m\times {\bf H}_n -{\bf E}_n\times {\bf H}_m\right] \cdot d{\bf S}.\label{scal2}
\end{equation}
with the permittivity $\varepsilon({\bf r})$ being a rotation symmetric even function of $z$, all integrals in $({\bf F}_1\cdot{\bf F}_3)$, $({\bf F}_1\cdot{\bf F}_4)$, $({\bf F}_2\cdot{\bf F}_3)$ and $({\bf F}_2\cdot{\bf F}_4)$ vanish by virtue of different parity of the field components with respect to $z$. 
For the states belonging to same degenerate pairs, e.g., for 1 and 2, the orthogonality follows from their mutual transformation according to  \eqref{rot1and2}.  Upon a rotation by $\pi$ about the $z$-axis, the scalar product analogue \eqref{scal2} is transformed as $({\bf F}_{1}\cdot{\bf F}_{2})\rightarrow-({\bf F}_{2}\cdot{\bf F}_{1})$ and, therefore, it should vanish due to its commutative property. Similarly, one proves that $({\bf F}_{3}\cdot{\bf F}_{4})=0$, and, therefore:
\begin{equation}
	({\bf F}_n\cdot{\bf F}_m)=0,\ n\neq m. \label{orthog}
\end{equation}

Note that the degeneracy determines ambiguity of choice of particular states within the pairs 1,2 and 3,4. Indeed, one can introduce a generalized 'rotation' with an arbitrary complex parameter $\varphi$ and, for example, combine another pair of states 3 and 4 as:
\begin{eqnarray}\label{rot}
	{\bf F}'_3=\cos\varphi{\bf F}_3+\sin\varphi{\bf F}_4,\\
	{\bf F}'_4=\cos\varphi{\bf F}_4-\sin\varphi{\bf F}_3,
\end{eqnarray} 
which then automatically remain normalized as in Equation~\eqref{norm}, orthogonal as in Equation~\eqref{orthog} and obey the transformation rules similar to \eqref{rot1and2} and \eqref{rot2and1}.

\medskip

Accordingly, using the states ${\bf F}_{1-4}$ as an orthonormal subspace basis, we expand a perturbed state ${\bf \tilde{F}}$ as: 
\begin{equation}
	\tilde{\bf F}=\sum_{m={1}}^4 a_{m}{\bf F}_m
\end{equation}
and the expansion coefficients can be obtained by applying the RSE \cite{Muljarov2010,Muljarov2018}. Note that although the RSE was originally developed to account for the permittivity variations occurring inside light scattering structures (where the set of basis states is complete), recently it was shown \cite{Sztranyovszky2023} that external perturbations can also be treated by the RSE up to the first order. As was demonstrated in \cite{Almousa2023,Almousa2024}, this is specifically applicable to variations of the environment. Here we treat a weakly refracting substrate as an environment perturbation,  truncate the RSE matrix equation to the four states of interest, and obtain a system of equations for the coefficients $a_{n}$ valid in first order:
\begin{equation}
	\omega_n a_{n}=\tilde\omega\sum_{m=1}^4(\delta_{nm}+V_{nm})a_{m}.\label{RSE}
\end{equation} 
The elements of the symmetric perturbation matrix are evaluated as: 
\begin{equation}
	V_{nm}=\int \delta\varepsilon({\bf r}) {\bf E}_n({\bf r})\cdot{\bf E}_m({\bf r}) dV,\label{Vjk}
\end{equation} 
where, in our case, the permittivity perturbation is expressed by the Heaviside step function $\theta$:
\begin{equation}
	\delta\varepsilon({\bf r})=\Delta\ \theta(-z-h/2), \ {\rm with}\ \Delta= (n_{\rm s}^2-1),	
\end{equation}
i.e., is constant everywhere below the PCS which has the total thickness $h$. 

\medskip

Among 10 independent elements of the symmetric matrix $V_{nm}$, those involving states from same degenerate pairs identically vanish, $V_{12}=V_{34}=0$, similarly to $({\bf F}_1\cdot{\bf F}_2)$ and $({\bf F}_3\cdot{\bf F}_4)$ as discussed above. Diagonal matrix elements within each pair are equal by rotation symmetry: $V_{11}=V_{22}$ and $V_{33}=V_{44}$. Also, some matrix elements involving states from different degenerate pairs are related as they transform into each other upon $\pi/2$ rotations: $V_{13}=V_{24}$, $V_{14}=-V_{23}$. Altogether, the perturbation matrix can be parameterized as:
\begin{equation}\label{Vnm}
	{\bf V}=\Delta
	\begin{bmatrix}
		v_1 & 0 & u & w\\
		0   & v_1 & -w& u\\
		u	& -w & v_3 & 0\\
		w	&  u & 0  & v_3
	\end{bmatrix},
\end{equation}
with the parameters 
\begin{eqnarray}
	v_1=\int_{z<-h/2} {\bf E}_1({\bf r})\cdot{\bf E}_1({\bf r}) dV, \label{v1}\ \ \ 
	v_3=\int_{z<-h/2} {\bf E}_3({\bf r})\cdot{\bf E}_3({\bf r}) dV, \label{v2}\\
	u=\int_{z<-h/2} {\bf E}_1({\bf r})\cdot{\bf E}_3({\bf r}) dV, \label{u}\ \ \ 
	w=\int_{z<-h/2} {\bf E}_1({\bf r})\cdot{\bf E}_4({\bf r}) dV. \label{w}
\end{eqnarray}
For further simplification, we use the ambiguity of choice of degenerate states and introduce a new pair of states 3 and 4 provided by a generalized rotation \eqref{rot} by an angle $\varphi_w=\arctan{w/u}$, which allows eliminating the parameter $w$. As a result, the matrix in the r.h.s. of Equation~\eqref{RSE} can be written as:
\begin{equation}\label{Vmatr}
	{\bf I}+{\bf V}=\left[1+\Delta\frac{v_1+v_3}{2}\right]\begin{bmatrix}
		1 & 0 & 0 & 0\\
		0   & 1 & 0& 0\\
		0	& 0 & 1 & 0\\
		0	&  0 & 0  & 1
	\end{bmatrix}+\Delta\begin{bmatrix}
		v & 0 & u & 0\\
		0   & v & 0& u\\
		u	& 0 & -v & 0\\
		0	&  u & 0  & -v
	\end{bmatrix},
\end{equation}
where $v=(v_1-v_3)/2$. In such representation, the substrate selectively mixes state 1 with state 3 and state 2 with state 4. 

\medskip

Solving Equation~\eqref{RSE}, one obtains a doublet of eigenfrequencies:
\begin{equation}
	\tilde\omega_\pm=\frac{1}{2(1-\tilde u^2)}\left[\tilde\omega_1+\tilde\omega_3\pm\sqrt{(\tilde\omega_1-\tilde\omega_3)^2+4\tilde u^2\tilde\omega_1\tilde\omega_3}\right],\label{tomegapm}
\end{equation}
where the frequency parameters $\tilde\omega_{1,3}= \omega_{1,3}(1+\Delta v_{1,3})^{-1}$ describe the direct spectral shift produced by the substrate, while the state splitting is governed by the parameter $\tilde u = u\Delta/\sqrt{(1+\Delta v_{1})(1+\Delta v_{3})}$.  

The branch $\tilde\omega_{+}$ corresponds to a pair of degenerate eigenstates expressed as:
\begin{equation}
	\tilde{\bf F}_{1}=\cos\phi{\bf F}_1+\sin\phi{\bf F}_3, \ 
	\tilde{\bf F}_{2}=\cos\phi{\bf F}_2+\sin\phi{\bf F}_4,\label{F12}
\end{equation}   
while $\tilde\omega_{-}$ is the eigenfrequency of another pair:
\begin{equation}
	\tilde{\bf F}_{3}=\cos\phi{\bf F}_3 -\sin\phi{\bf F}_1,\ 
	\tilde{\bf F}_{4}=\cos\phi{\bf F}_4-\sin\phi{\bf F}_2\label{F34}.
\end{equation}   
The angle parameter $\phi$ characterizes the state mixing, and it is determined by: 
\begin{equation}
	\cot2\phi=\frac{v}{u}+\frac{\omega_3-\omega_1}{2\Delta u\tilde\omega_-}.\label{cotphi}
\end{equation}
which shows that the substrate induces a strong hybridization in the vicinity of intersection of spectral branches of unperturbed states, as long as  $|\omega_3-\omega_1|\lesssim{\Delta |u|\omega_{1,3}}$. Otherwise, when the second term in Equation~\eqref{cotphi} is much larger than unity, $\phi\simeq0,\pi/2$, and the perturbed states (\ref{F12}--\ref{F34}) are approximately equal to the unperturbed ones. Note that the orthogonality and norm of the perturbed states are explicitly ensured by those of the unperturbed states \eqref{orthog}.
As shown in SI Section~S2, remarkable specific conditions can be derived from the fact that the considered eigenstates possess high quality factors, i.e., are close to BICs. Then the parameters $v$ and $u$ are real, and the mixing angle $\phi$ stays also almost real as a substantial imaginary contribution to Equation~\eqref{cotphi} arises from the unequal decay rates of unperturbed states,  and it becomes substantial only for sufficiently small $\Delta$.  

\section{Optical chirality realized with hybrid states}\label{sec:chiral}

Most generally, the light transmission and reflection by a metasurface can be described as an S-matrix problem. For subwavelength metasurface periodicity, the diffraction is absent, and the problem includes in total 4 input and 4 output ports for 2 metasurface sides and for 2 independent polarizations for each propagation direction. To analyze optical chirality at normal incidence, one introduces circular polarizations with the complex electric field amplitude along the unit vectors ${\bf e}_\pm=({\bf e}_x\pm i{\bf e}_y)/\sqrt{2}$. Here ${\bf e}_+$ describes LCP and  RCP waves propagating, respectively, along and against the $z$-axis, and the opposite is true for  ${\bf e}_-$.

\medskip

The resonant-state scattering theory allows expressing the S-matrix elements by a pole expansion \cite{Weiss2018}:
\begin{equation}\label{Smat}
	S_{\rm fi} = C_{\rm fi} +\sum_{n}\frac{\mathcal{R}_{n,\rm fi}}{\omega-\omega_n},
\end{equation}  
where the indexes ${\rm f}$ and ${\rm i}$ denote the final and initial states of light, i.e., the corresponding input and output channels of the S-matrix problem. In the following, we denote by the indexes $\rm R$ and $\rm L$ RCP and LCP waves incoming/outgoing to/from one metasurfaces side, and by the primed ones,  $\rm R'$ and $\rm L'$, those on the other side.  
$C_{\rm fi}$ is an element of the background non-resonant S-matrix, $\omega_n$ is a complex eigenfrequency of state $n$, and $\mathcal{R}_{n,\rm fi}$ is an S-matrix residue describing contribution of state $n$ to the scattering from channel $\rm i$ to channel $\rm f$. 

\medskip

As discussed in SI Section~S3, very different approaches, such as  the phenomenological coupled-mode theory \cite{Fan2003} and the resonant S-matrix theory \cite{Weiss2018}, yield the residues in the form  $\mathcal{R}_{n,\rm fi}=im_{n,\rm f}m_{n,\rm i}$, i.e.,  expressed as products of the parameters $m_{n,\rm f}$ and $m_{n,\rm i}$ quantitatively characterizing the coupling of state $n$ to the corresponding channels. For a PCS in symmetric vacuum environment, transmitting normally incident plane waves, the coupling parameters are expressed by the overlap integrals of the eigenstate electric field ${\bf E}_{n}$ with electric fields of RCP and LCP plane waves propagating in the corresponding direction and having the same frequency $\omega_n$:
\begin{equation}
	{m}_{n, \rm R}=i\frac{\omega_n}{\sqrt{2Ac}}\ \int_{ V}[\varepsilon({\bf r})-1]({\bf {E}}_{n}\cdot {\bf e}_{+})e^{-i\omega_nz/c} dV,\
	{m}_{n, \rm L}=i\frac{\omega_n}{\sqrt{2Ac}}\ \int_{ V}[\varepsilon({\bf r})-1]({\bf {E}}_{n}\cdot {\bf e}_{-})e^{-i\omega_nz/c} dV,\label{mRL}
\end{equation}
for the top (as in Figure~\ref{fig:1}a) side, and 
\begin{equation}
	{m}_{n, \rm R'}=i\frac{\omega_n}{\sqrt{2Ac}}\ \int_{ V}[\varepsilon({\bf r})-1]({\bf {E}}_{n}\cdot {\bf e}_{-})e^{i\omega_nz/c} dV,\
	{m}_{n, \rm L'}=i\frac{\omega_n}{\sqrt{2Ac}}\ \int_{ V}[\varepsilon({\bf r})-1]({\bf {E}}_{n}\cdot {\bf e}_{+})e^{i\omega_nz/c} dV,\label{m1RL}
\end{equation}
for the opposite bottom side.
Here the integrals are taken over the volume $V$ of one PCS unit cell, $A$ is the unit cell area, and the factor $[\varepsilon({\bf r})-1]$ reduces the integration to the volume occupied by the high refractive index dielectric forming the PCS.

\medskip
Depending on the eigenstate parity, the parameters of coupling on different sides are related to each other as ${m}_{n, \rm R}=\pm {m}_{n, \rm L'}$ and ${m}_{n, \rm L}=\pm {m}_{n, \rm R'}$. Supposing that the non-resonant background transmission is achiral and isotropic, we characterize it by a single coefficient $\tau$ and obtain identical S-matrix elements describing the co-polarized transmission: 
\begin{equation}
	S_{\rm L'L}=\tau-\sum_n \frac{{m}_{nL'}{m}_{nL}}{i(\omega-\omega_n)}= S_{\rm R'R}= \tau-\sum_n \frac{{m}_{nR'}{m}_{nR}}{i(\omega-\omega_n)}.
	\label{SLLRR}
\end{equation}
This generally excludes the possibility of the transmission CD of a PCS in symmetric environment.

\medskip

Eigenstate degeneracy due to the PCS rotation symmetry allows impose additional conditions on pairs of PCS eigenstates. Thus  one can choose between different linear combinations of states 1 and 2. For clarity, we set state 1 to be linearly polarized along the $y$-axis, which determines state 2 to be linearly polarized along the $x$-axis. States 3 and 4 are then to be chosen accordingly to exclude the cross coupling with $w=0$. As is discussed in SI Section~S4, this defines the whole set of modes ${\bf F}_{1-4}$ with all relative phases fixed by the norm \eqref{norm} and all parameters of coupling to linearly polarized waves expressed by a few constants: two strengths of coupling  $M_{1,3}$ and the angle $\psi$ between the far-field linear polarization of states 1 and 3 (or, equivalently, states 2 and 4) shown in Figure~\ref{fig:1}(c). The coupling  parameters then take simple form:
\begin{gather}\label{m12}
	{m}_{1, \rm R}={m}_{1, \rm R'}=-{m}_{1, \rm L}=-{m}_{1, \rm L'}=\frac{i}{\sqrt{2}}M_1,\\
	{m}_{2, \rm R}=-{m}_{2, \rm R'}={m}_{2, \rm L}=-{m}_{2, \rm L'}=\frac{1}{\sqrt{2}}M_1,\\
	{m}_{3, \rm R}={m}_{3, \rm L'}=\frac{i}{\sqrt{2}}M_3e^{-i\psi},\
	{m}_{3, \rm L}={m}_{3, \rm R'}=-\frac{i}{\sqrt{2}}M_3e^{i\psi},\\
	{m}_{4, \rm R}={m}_{4, \rm L'}=\frac{1}{\sqrt{2}}M_3e^{-i\psi},\
	{m}_{4, \rm L}={m}_{4, \rm R'}=\frac{1}{\sqrt{2}}M_3e^{i\psi}.
	\label{m34}
\end{gather}

\medskip

Naturally, introducing a substrate, alters all parameters entering the S-matrix \eqref{Smat}, however, to different extent. Apart from the already considered transformation of the eigenfrequencies \eqref{tomegapm}, the eigenstate fields become mixed in linear combinations \eqref{F12} and \eqref{F34}, and their hybridized electric field distributions $\tilde{\bf E}_n$ are now to be integrated to obtain the corresponding hybrid coupling parameters $\tilde{m}$. This gives rise to the major effect driven by the substrate, as the states are strongly mixed when the angle  $\phi$ is different from a multiple of $\pi/2$.

\medskip

Other changes caused by weakly refracting substrates are less significant. Thus, instead of incoming and outgoing plane waves as in \eqref{mRL} and \eqref{m1RL}, one should formally substitute  their linear combinations including contributions due to weak reflection and non-unitary transmission by the substrate. Also, similarly small corrections have to be introduced to the state normalization. In order to keep the consideration reasonably simple, in the first  order of RSE perturbation theory, we neglect such weak corrections and focus on qualitative changes caused by strong eigenstate hybridization.

\medskip

Therefore, in the first approximation, the coupling parameters of the hybridized states \eqref{F12} and \eqref{F34} can be expressed as:
\begin{eqnarray}
	\tilde m_{1R}= i\tilde m_{2R}=\frac{i}{\sqrt{2}}(M_1  \cos\phi +  M_3   e^{-i\psi} \sin\phi),\label{mR2}\\
	\tilde m_{1R'}=-i\tilde m_{2R'}=\frac{i}{\sqrt{2}} (M_1 \cos\phi - M_3   e^{i\psi} 	\sin\phi),\label{mR21}\\
	\tilde m_{1L}=-i\tilde m_{2L}= \frac{-i}{\sqrt{2}}(M_1 \cos\phi  + M_3   e^{i\psi} \sin\phi),\label{mL2}\\
	\tilde m_{1L'}=i\tilde m'_{2L}=\frac{-i}{\sqrt{2}}(M_1  \cos\phi -  M_3   e^{-i\psi} \sin\phi),\label{mL21}\\
	\tilde m_{3R} =i\tilde m_{4R}=\frac{-i}{\sqrt{2}}(M_1 \sin\phi - M_3  e^{-i\psi} \cos\phi),\label{mR4}\\
	\tilde m_{3R'} =-i\tilde m_{4R'}= \frac{-i}{\sqrt{2}}(M_1 \sin\phi + M_3   e^{i\psi} \cos\phi),\label{mR41}\\
	\tilde m_{3L} =-i\tilde m_{4L}=\frac{i}{\sqrt{2}} (M_1 \sin\phi - M_3   e^{i\psi} \cos\phi),\label{mL4}\\
	\tilde m_{3L'} =i\tilde m_{4L'}=\frac{i}{\sqrt{2}}(M_1 \sin\phi + M_3  e^{-i\psi} \cos\phi).\label{mL41}	
\end{eqnarray}

These parameters together with the eigenfrequencies \eqref{tomegapm} determine the S-matrix components describing the co-polarized transmission by the PCS on a substrate:
\begin{gather}
	\tilde{S}_{\rm L'L}=\tau-2 \frac{\tilde{m}_{1L'}\tilde{m}_{1L}}{i(\omega-\tilde\omega_+)}-2 \frac{\tilde{m}_{3L'}\tilde{m}_{3L}}{i(\omega-\tilde\omega_-)}
	\label{tSLL}\\
	\tilde{S}_{\rm R'R}=\tau-2 \frac{\tilde{m}_{1R'}\tilde{m}_{1R}}{i(\omega-\tilde\omega_+)}-2 \frac{\tilde{m}_{3R'}\tilde{m}_{3R}}{i(\omega-\tilde\omega_-)}
	\label{tSRR}
\end{gather}
It is convenient to quantify the contribution of a hybrid eigenstate to the metasurface co-polarized transmittance $T_{LL}=|\tilde{S}_{\rm L'L}|^2$ and  $T_{RR}=|\tilde{S}_{\rm R'R}|^2$ by a state CD defined as \cite{Toftul2024}:
\begin{equation}
	CD_n=\frac{|\tilde{m}_{nR}\tilde{m}_{nR'}|^2-|\tilde{m}_{nL}\tilde{m}_{nL'}|^2} {|\tilde{m}_{nR}\tilde{m}_{nR'}|^2+|\tilde{m}_{nL}\tilde{m}_{nL'}|^2},\label{CDn}
\end{equation}
Maximum optical chirality is achieved when $|\tau|=1$ and a spectrally well separated hybrid eigenstate possesses $CD_n=\pm 1$. Indeed, then one of the transmission coefficients \eqref{tSLL} or \eqref{tSRR} remains equal to 1 being unaffected by this state, while the other one experiences a resonant dip which can reach zero upon fulfilling the critical coupling condition \cite{Gorkunov2020}. For $CD_n=\pm 1$, either of the coupling parameters has to vanish, i.e., the state has to be a chiral quasi-BIC fully isolated from LCP or RCP waves on either metasurface side~\cite{Gorkunov2020}.
Note that for the coupling parameters (\ref{m12}--\ref{m34}), the state $CD_n$ vanishes identically when $\psi=0,\pi$, that is when the linear polarization of the coupled states is identical in the far field, or when $\phi=0,\pi$, which means the absence of state hybridization.  

\medskip

It is not difficult to reveal the analytical conditions of maximum chirality of hybrid states. For example, to establish a maximum chiral LCP-polarized transmission resonance at frequency $\tilde\omega_+$, one should eliminate the corresponding coupling parameters either as $\tilde m_{2R}= -i\tilde m_{1R}=0$ or as $\tilde m_{2R'}=i\tilde m'_{1R'}=0$, i.e., to fulfill either of the conditions:
\begin{equation}
	\frac{M_3}{M_1}=-e^{i\psi}\cot\phi , \ {\rm or}\  \frac{M_3}{M_1}=e^{-i\psi}\cot\phi, \label{LCP+}.
\end{equation} 
To ensure simultaneous maximum chirality of the other transmission resonance at $\tilde\omega_-$, one of the coupling coefficients of modes 3 and 4 has to vanish. It is impossible to establish another LCP-polarized resonance by fulfilling $\tilde m_{4R} =-i\tilde m_{3R}=0$ or  $\tilde m_{4R'}=i\tilde m_{3R'}=0$. For real $\phi$, this would set $\psi=0,\pi$ and eliminate the chirality. 

\medskip

At the same time, establishing an RCP-polarized resonance at $\tilde\omega_-$ requires setting either $\tilde m_{4L} =i\tilde m_{3L}=0$ or $\tilde m_{4L'} =-i\tilde m_{3L'}=0$, i.e., fulfilling one of the conditions:
\begin{equation}
	\frac{M_3}{M_1}=e^{-i\psi}\tan\phi , \ {\rm or}\  \frac{M_3}{M_1}=-e^{i\psi}\tan\phi, \label{RCP-}
\end{equation} 
which are compatible with Equation~\eqref{LCP+}, when, for instance, 
\begin{equation}
	\phi=\pi/4\ {\ \rm and}\ \ \frac{M_3}{M_1}=\pm e^{\mp i\psi}.\label{ultima}
\end{equation}
The physical meaning of these conditions is clear: the substrate mixes unperturbed states of different parity in the most equal proportion (with $\phi=\frac{\pi}{4}$) and these states are equally strongly (as $|M_1|=|M_3|$) coupled to free-space waves with the phase shift corresponding to the 'twist angle' $\psi$.

\begin{figure}
	\centering
	\includegraphics[width=\linewidth]{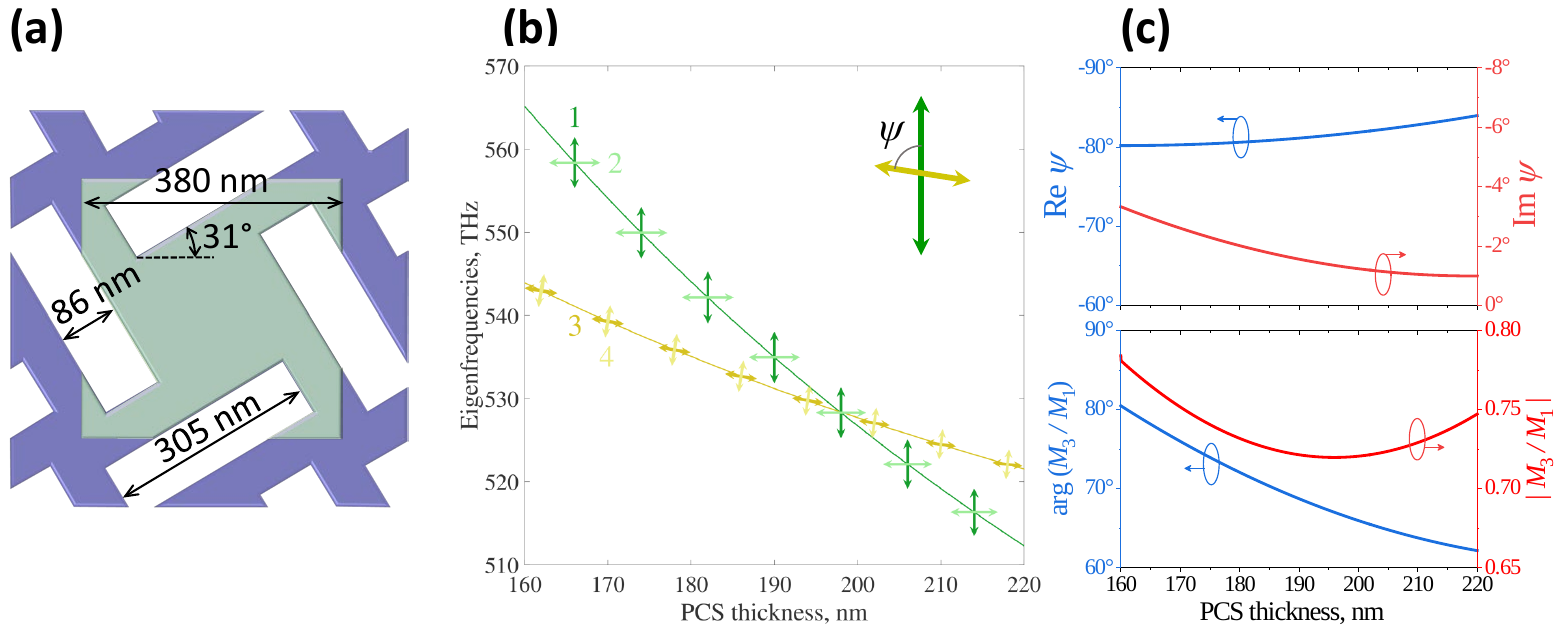}
	\caption{Achiral PCS in symmetric environment and its eigenstates. (a) Fragment of the PCS structure with a square unit cell highlighted and the key dimensions indicated. (b) Real part of eigenfrequencies of degenerate pairs of odd (1 and 2) and even (3 and 4) states as functions of PCS thickness. The far-field polarization is indicated and the arrow size is inversely proportional to the state quality factor. (c) Real and imaginary parts of the angle $\psi$ and the phase and magnitude of the the ratio $M_3/M_1$ as functions of the PCS thickness.}\label{fig:fig2} 
\end{figure}	
	
\section{Examples of substrate-induced maximum chirality}\label{sec:comsol}
	
To exemplify the general principle of substrate-driven optical chirality, we model in COMSOL Multiphysics a PCS build as a 4-fold rotation-symmetric array of rectangular holes in a layer of dielectric material with a refractive index $n_{\rm PCS}=2.2+0.014 i$. The moderately high real part of $n_{\rm PCS}$ is close to that of widely studied optoelectronic semiconductors, such as GaN or various metal halide perovskites in wavelength ranges close to their excitonic bands. Small imaginary part is generally inherent to such materials, and for the optical chirality here it is crucial as the absorption circular dichroism causes all chiral optical effects in $C_4$ rotation-symmetric structures. The particular small value $0.014i$ is chosen empirically to ensure closeness to the critical coupling condition when the absorption and radiation decay rates of the states are equal. Fulfilling this condition allows establishing maximum chiral resonances, when waves of a particular circular polarization are fully blocked while those of the opposite polarization freely pass through the metasurface \cite{Gorkunov2020}.  

\medskip

\begin{figure}
	\centering
	\includegraphics[width=0.9\textwidth]{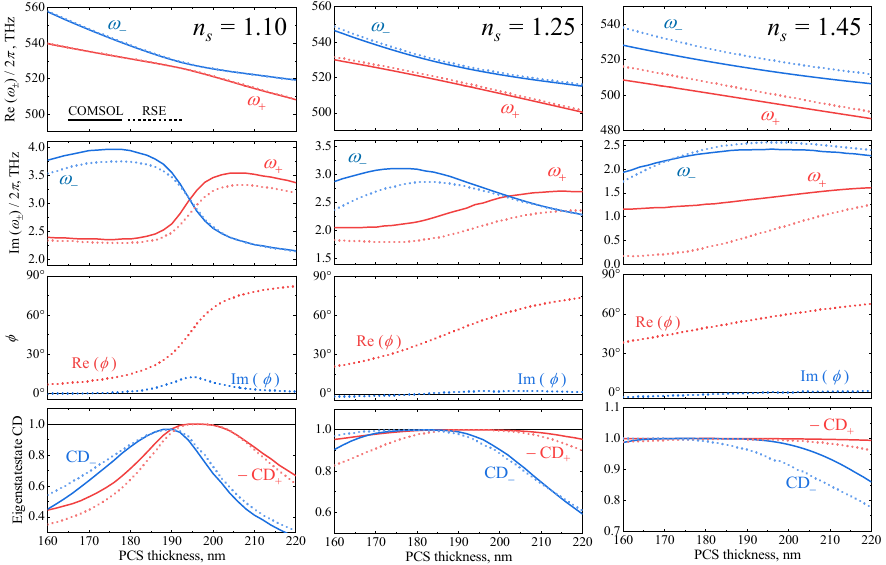}
	\caption{Dependence of the real (first row) and imaginary (second row) parts of eigenfrequencies $\tilde\omega_{\pm}$ obtained by COMSOL Multiphysics (solid) and the RSE theory via Equation~\eqref{tomegapm} (dashed) as functions of the PCS thickness. The third row shows the corresponding values of mixing angle $\phi$ obtained from Equation~\eqref{cotphi}. The fourth row depicts the state CDs evaluated by the coupling parameters obtained by simulations (solid) and by the RSE-based theory via Equations~(\ref{mR2}--\ref{mL41}) (dashed). Three columns correspond to three indicated substrate refractive indexes. All PCS parameters are as in Figure~\ref{fig:fig2}.}\label{fig:CDs} 
\end{figure}

\medskip

We consider a PCS built as a rectangular hole pattern shown in Figure~\ref{fig:fig2}(a). For definiteness, we pick an operational wavelength range between 550~nm and 650~nm, and choose the square lattice period of 380~nm to avoid diffraction into a glass-like substrate with relatively high $n_{\rm s}=1.45$. The centers of the holes are located at the middle points of the sides of square unit cell, and their size and rotation are obtained by the COMSOL optimization to ensure close-to-unity eigenstate $CD_n$ when the PCS is placed on a glass-like substrate. The PCS thickness is a convenient parameter allowing us to control the spectra of eigenstates, and we use it to establish small difference between the eigenfrequencies of states of different parity in the absence of a substrate.

\medskip

Employing COMSOL eigenstate solver, we obtain 4 relevant states and normalize them according to Equation~\eqref{norm}. As one can see in Figure~\ref{fig:fig2}(b), the two spectral branches of degenerate pairs of odd (1 and 2) and even (3 and 4) states intersect at about a thickness of 200~nm. Their polarization is evaluated by analyzing the coupling parameters \eqref{mRL} and \eqref{m1RL}.  According to the general routine described in SI Section~S4, we choose states 1 and 2 with the far fields linearly polarized in the $y$ and $x$ directions respectively, and transform states 3 and 4 to eliminate the parameter $w$ of coupling of states 1 and 4 (equivalently, 2 and 3). We obtain that states 3 and 4 are also linearly polarized and vividly see a large twist angle $\psi$ between the polarization of states 1 and 3 (equivalently 2 and 4), which is almost purely real and stays close to $-\pi/2$, see Figure~\ref{fig:fig2}(c). At the same time, the ratio of coupling strengths $M_3/M_1$ is close to purely imaginary thus approaching the condition \eqref{ultima}, although its absolute value remains somewhat lower than unity. This illustrates a natural limitation of chiral PCS design. One can adjust the parameters to ensure crossing of the real parts of frequencies, however, their imaginary parts depending on the radiative decay rates of the states and the strength of their coupling to free-space waves still remain somewhat different for the states of different parity.

\begin{figure}
	\centering\includegraphics[width=\linewidth]{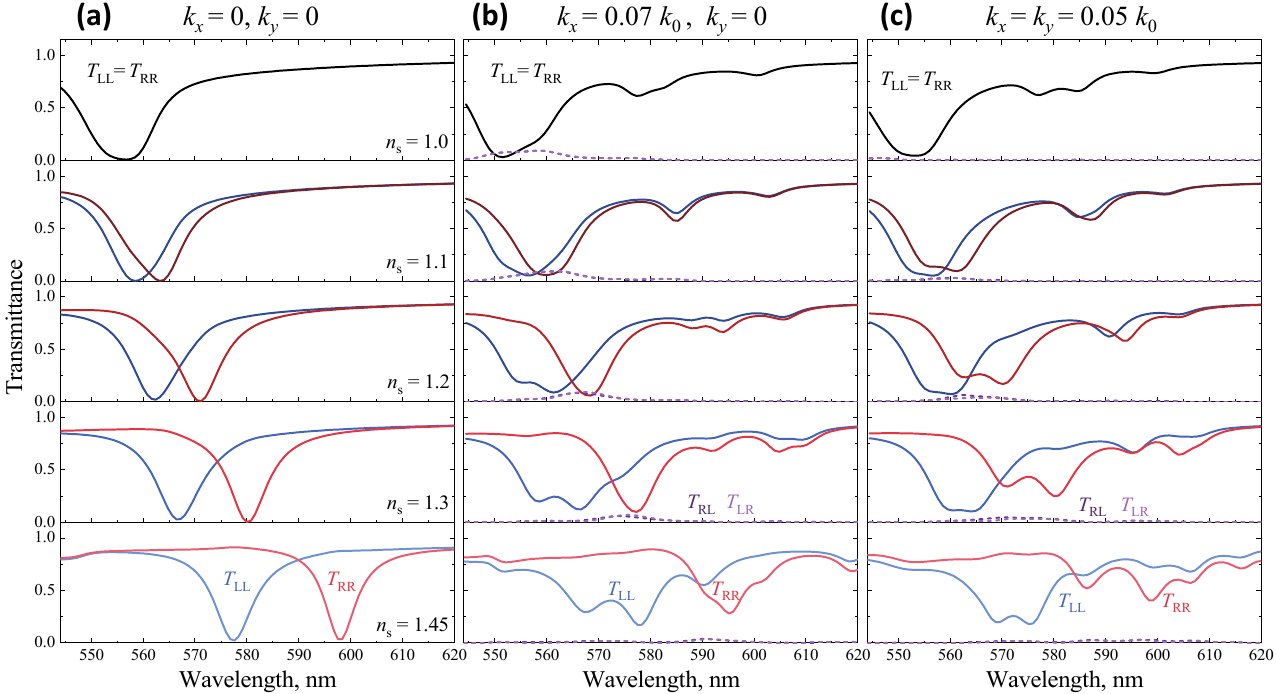} \caption{Spectra of co-polarized ($T_{LL}$ and $T_{RR}$, solid lines) and cross-polarized ($T_{LR}$ and $T_{RL}$, dashed lines) transmission of circularly polarized waves by PCS with no substrate (black lines in the top row) and PCS on substrates with various indicated values of the refractive index $n_{\rm s}$ in all other rows. The PCS thickness is 180~nm and the material refractive index is $n_{\rm PCS}=2.2+0.014 i$. The cases of normal (a) and two oblique incidences (b) and (c) are shown. The wavevector components in terms of the vacuum wavenumber $k_0=\omega/c$ are indicated on the top.}\label{fig:spectra} 
\end{figure}

\medskip

Next, we model the eigenstates of PCS on substrates with different $n_{\rm s}$ and compare their properties with those predicted by the first-order RSE based on the properties of unperturbed states. Representative results are shown in Figure~\ref{fig:CDs}. As the substrate lifts the PCS environment symmetry, the state coupling transforms the spectral crossing into avoided crossing. Using unperturbed eigenfrequencies $\omega_1$ and $\omega_3$ we calculate the eigenfrequency branches predicted by the RSE theory according to Eq.~\eqref{tomegapm} for different values of the perturbation strength $\Delta$ determined by $n_{\rm s}$. Comparing them with the accurate numerically obtained spectra (see solid and dashed curves in the first two rows in Figure~\ref{fig:CDs}) allows us to identify the limits of validity of the first-order RSE. The perturbative approach remains reasonably accurate for the real frequency parts. For the much smaller imaginary parts, however, the relative error becomes substantial already at $n_{\rm s}=1.25$, which suggests that one should be cautious when using the RSE for predicting quality factors. 

\medskip
	
As one can see in Figure~\ref{fig:CDs}, in agreement with our expectations, the mixing angle $\phi$ is close to $\pi/4$ near the spectra intersection point. It becomes almost purely real for the substrates with  $n_{\rm s}>1.2$. Its imaginary part for $n_{\rm s}=1.1$ corresponds to a stronger contribution of the second term in Equation~\eqref{cotphi} for smaller $\Delta$. Also in-line with our expectations, the state CDs approach their extreme $\pm1$ values near the crossing point when $\phi\approx\pi/4$, i.e., when odd and even states are strongly mixed. Although this happens not exactly at the same thickness when $n_{\rm s}=1.1$, for stronger refracting substrates, $n_{\rm s}>1.2$, the hybridization occurs in a wider range of thicknesses, and both CDs remain remarkably close to $\pm1$ there. Again, the perturbation RSE theory reproduces state CDs relatively accurately well up to $n_{\rm s}\approx1.2$, but quantitative deviations become obvious at larger $n_{\rm s}$. 

\begin{figure}
	\centering\includegraphics[width=\linewidth]{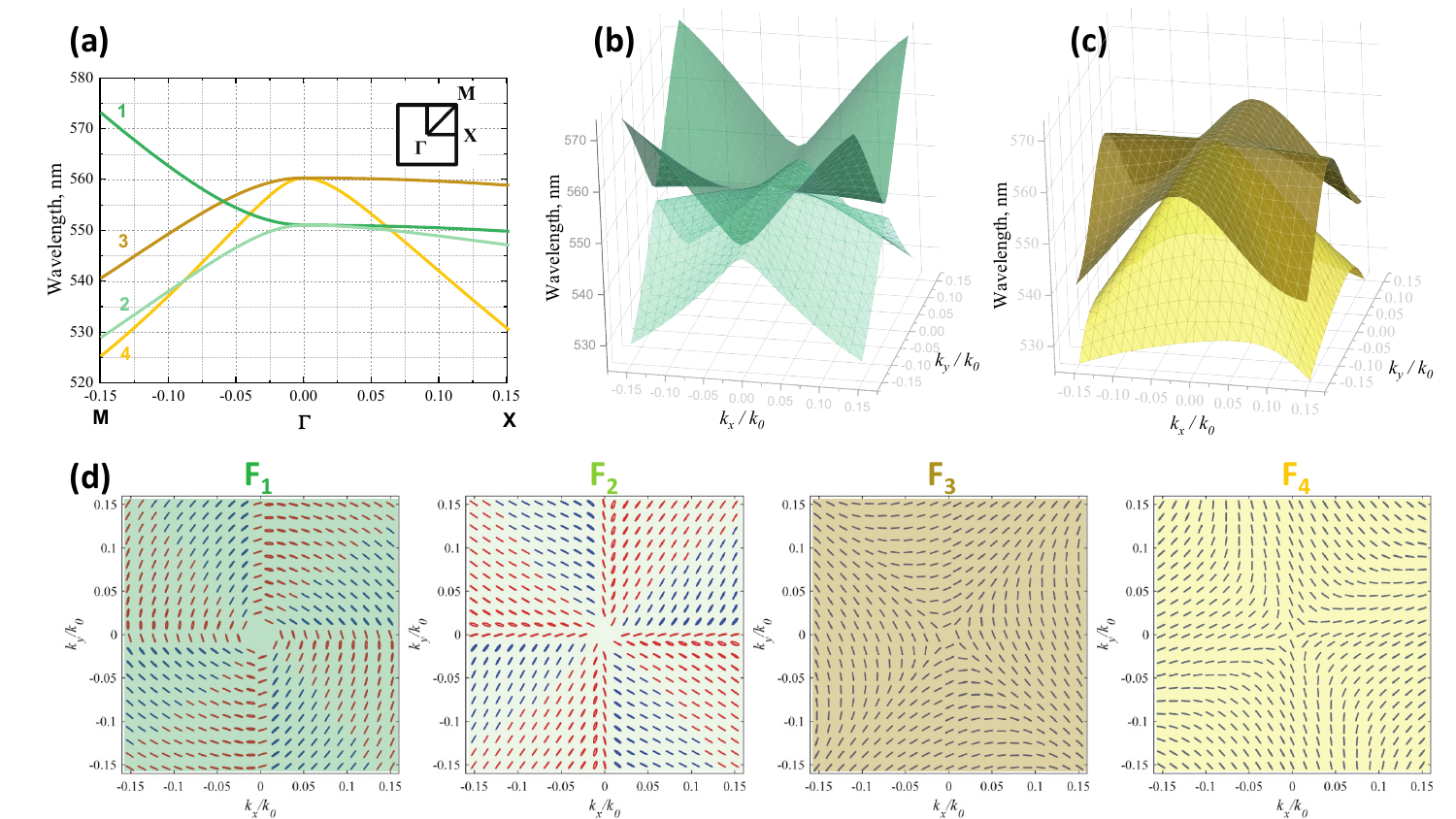} \caption{Eigenstates of achiral PCS in symmetric environment ($n_{\rm s}=1$) in the momentum space around the $\Gamma$-point. Band structure including pairs of states 1,2 and 3,4 degenerate at the $\Gamma$-point (a). Wavelengths corresponding to the real parts of eigenfrequencies of states 1,2 (b) 3,4 (c) as functions of the in-plane wavenumbers. Polarization maps of the far-field plane-wave asymptotics of all four states (d). The PCS dimensions and material parameters are as in Figure~\ref{fig:spectra}.} \label{fig:oblique_achiral} 
\end{figure}

\medskip

Next, by solving the transmission problem in COMSOL Multiphysics, we evaluate the spectra of transmission of circularly polarized waves and verify that the symmetry-breaking effect of the substrate transforms the achiral PCS transmission resonance into a pair of maximum-chiral resonances of opposite handedness. As is shown in Figure~\ref{fig:spectra}, strong chirality is achieved already for $n_{\rm s}=1.2$. Remarkably, it is manifested not only for normally incident light (as in Figure~\ref{fig:spectra}(a)) but also for oblique directions, see Figures~\ref{fig:spectra}(b) and \ref{fig:spectra}(c) showing the transmission of waves tilted by about $4^\circ$ in the directions along the side and along the diagonal of the PCS unit cell, respectively. One can see that the tilting does not result in considerable cross--polarized transmission which is forbidden by rotation symmetry for normal incidence. At the same time, the chiral co-polarized transmission resonances are noticeably affected as they become split and  shallower.

\medskip

Interestingly, the larger-wavelength $T_{RR}$ resonance is stronger affected by the tilting, while the $T_{LL}$ resonance although broadens but shows a remarkable stability retaining a considerable CD for all tilting directions. For example, for the PCS upon a glass-like substrate with $n_{\rm s}=1.45$ (see the last row in Figure~\ref{fig:spectra}), the LCP transmission resonance for the normal incidence occurs at a wavelength of 578~nm with the transmittance difference $T_{RR}-T_{LL}=0.89$. For the oblique incidence, the LCP resonance with such difference in the range 0.66--0.71 occurs at wavelengths of 576--578 nm.  

\medskip

\begin{figure}
	\centering\includegraphics[width=\linewidth]{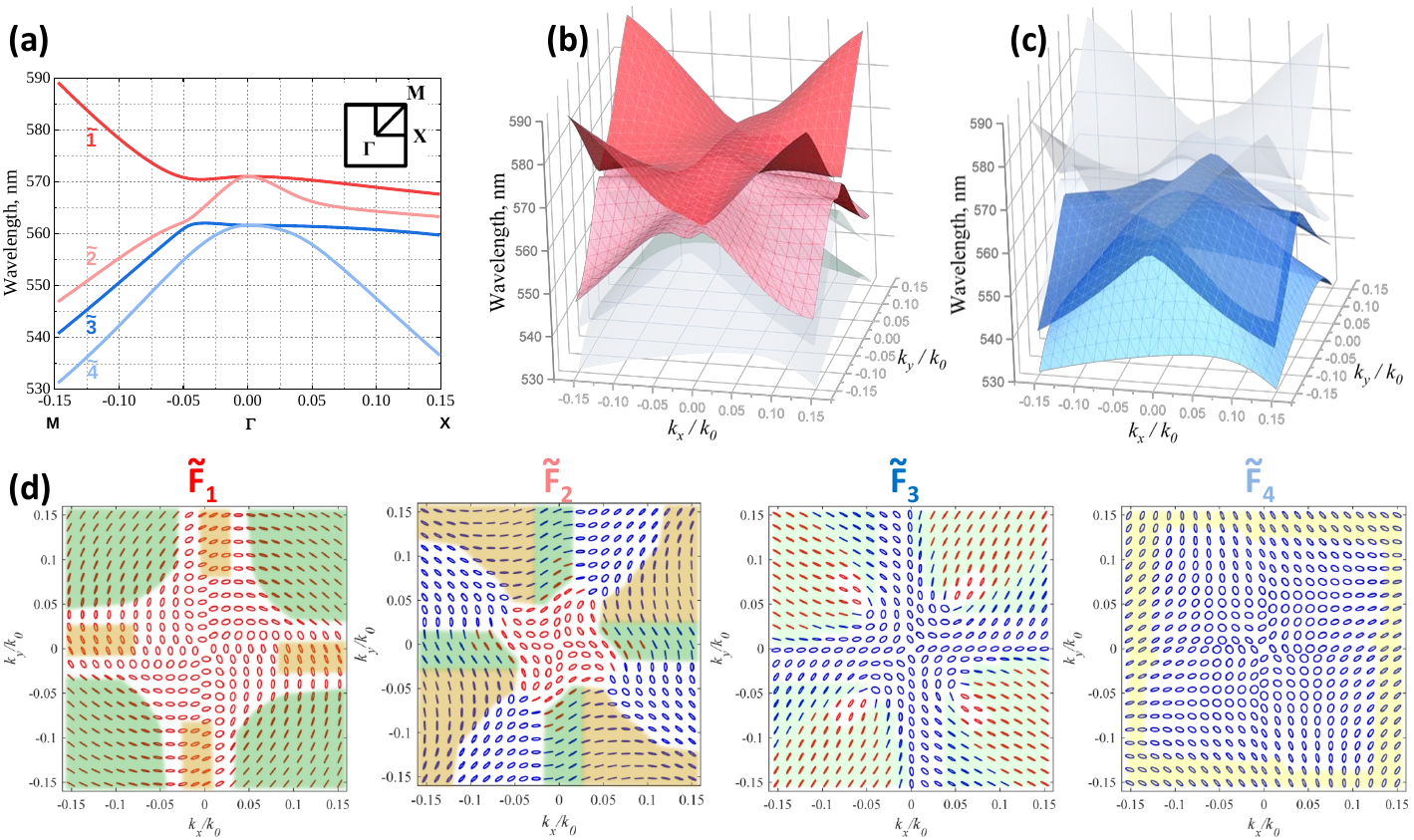} \caption{Eigenstates of the chiral PCS on a substrate with $n_{\rm s}=1.2$ around the $\Gamma$-point in the momentum space. Band structure including pairs of states $\tilde 1$,$\tilde 2$ and $\tilde 3$,$\tilde 4$ degenerate at the $\Gamma$-point (a). Wavelengths corresponding to the real parts of eigenfrequencies of states $\tilde 1$,$\tilde 2$ (b) $\tilde 3$,$\tilde 4$ (c) as functions of the in-plane wavenumbers.  Polarization maps of the far-field plane-wave asymptotics of all four states (d) with the fragments similar to the polarization maps of the unperturbed states as in Figure~\ref{fig:oblique_achiral}(d) indicated by the background colors same as in Figure~\ref{fig:oblique_achiral}(d). The white areas correspond to strong state coupling resulting in larger far-field ellipticity. The PCS dimensions and material parameters are as in Figure~\ref{fig:spectra}.} \label{fig:oblique_chiral} 
\end{figure}

To clarify such peculiar behavior of the transmission chirality, we study the metasurface eigenstate band structure and the far field  polarization in oblique directions. Those for the achiral PCS in symmetric environment are shown in Figure~~\ref{fig:oblique_achiral}, while typical eigenstate properties of the PCS upon a substrate can be seen in Figure~\ref{fig:oblique_chiral} for $n_{\rm s}=1.2$. 

\medskip

According to Figure~~\ref{fig:oblique_achiral}(a), the relevant branches of eigenstates of the same parity coincide only at the $\Gamma$-point and noticeably depart from each other for nonzero in-plane component of the wavevector. The surfaces in Figures~\ref{fig:oblique_achiral}(c) and \ref{fig:oblique_achiral}(d) show how the eigenfrequencies of states 1,2 remain closer in the direction of the X-point but strongly differ in most of other directions. For states 3,4 such specific  direction is towards the M-point. 
Different parity excludes coupling of states 1,2 with states 3,4, and their branches intersect.

\medskip

As is typical for eigenstates of a PCS in symmetric environment \cite{Zhen2014,Hsu2017}, their far field is almost perfectly linearly polarized, as one can see in Figure~\ref{fig:oblique_achiral}(d). At the $\Gamma$-point, their polarization is undefined, as one can chose arbitrary linear combination of the degenerate states. This allows the $\Gamma$-point to carry a unit topological charge. 

\medskip

The substrate breaking the out-of-plane mirror symmetry allows the states of different parity to hybridize. Then all spectral crossings seen in Figure~\ref{fig:oblique_achiral}(a) become avoided crossings seen in Figure~\ref{fig:oblique_chiral}(a). As a result, the eigenfrequency surfaces rearrange as shown in Figure~\ref{fig:oblique_chiral}(b).
The corresponding eigenstate polarization maps appear to be similar in both directions (into the air and into the substrate), and are shown in Figure~\ref{fig:oblique_chiral}(d). One can see that certain linearly or almost linearly polarized parts of these maps resemble fragments of maps of unperturbed states in Figure~\ref{fig:oblique_achiral}(d). This is naturally explained by the fact that within such fragments the eigenfrequencies are far from avoided crossings, the states are not substantially hybridized and remain close to the unperturbed ones.

\medskip

Importantly for the optical chirality, hybrid states occurring everywhere near the avoided crossings in the momentum space possess pronounced ellipticity, i.e., carry helicity. In all 4 polarization maps in Figure~\ref{fig:oblique_chiral}(d), one can see that a considerable part of the momentum space around the $\Gamma$-point   is occupied by such states. This explains why the PCS upon a substrate retains optical chirality in oblique directions. The stability of eigenfrequency of a particular chiral resonance is explained by the flatness of the eigenfrequency surface of state $\tilde 3$ vividly seen in Figure~\ref{fig:oblique_chiral}(a)

\section{Summary and conclusion}\label{sec:concl}

We have revealed that close-to-maximum optical chirality can be achieved with a planar PCS when its mirror symmetry is broken by a transparent substrate with a relatively low value of the refractive index. The key preconditions are:
\begin{itemize}
	\item[-] the eigenfrequencies of PCS eigenstates of different parity are close enough so that the states strongly hybridize by a coupling through the substrate, see Equations~(\ref{F12}--\ref{cotphi}); 
	\item[-] the in-plane PCS asymmetry induces considerable twisting angle $\psi$ of the eigenstate far-field relatively to its near-field to ensure relations as \eqref{RCP-};
	\item[-] for rotation-symmetric designs, weak absorption of light in the PCS material should ensure the eigenstate critical coupling condition so that its absorption and radiation decay rates are equal, as it is illustrated by the simulations in Section~\ref{sec:comsol}.
\end{itemize}

In the considered case, the rotation symmetry substantially simplifies identifying the effects of geometric chirality, as it excludes all other types of polarization effects for the transmission of normally incident waves. One can find in the literature many examples of planar metastructures of lower symmetry or totally lacking point symmetry elements, which, being  placed upon substrates, exhibit asymmetric co- and cross-polarized transmission CD, see, e.g. Refs~\cite{Kim2021a,Zong2022, Liu2023, Luo2023, Gryb2023, Koshelev2023}. In such cases, one cannot unambiguously identify the effects of the out-of-plane mirror symmetry breaking. Note that achiral PCSs of lower rotation symmetry can also perform as helicity preserving mirrors in purely symmetric environment and selectively reflect waves of one circular polarization while transmitting others inverting the helicity \cite{Semnani2020,Voronin2022}.

\medskip

In the general context of metasurface design development, 
this is another example of great variety of peculiar effects arising in relatively simple configurations supporting photonic states of different spatial parity, known also as TE- and TM-like modes in waveguides, or electric and magnetic states in Mie-type resonators. Even in the absence of intrinsic interaction, such states, by virtue of interference, can give rise to the generalized Kerker effect facilitating precise control over the transmitted light phase \cite{Decker2015}. In PCSs, the coexistence of such states is essential for helicity-preserving mirrors \cite{Semnani2020,Voronin2022, Duan2023}. Coupling between them can occur through the chiral environment which can be useful for enhanced sensing of weak natural chirality \cite{Chen2021}. Very recently, such states strongly coupled up to an exceptional point have been predicted to eventually acquire considerable chirality of reflections in particular directions \cite{Su2024}. 

\medskip

In conclusion, we have formulated the general principles for a design of planar photonic structures including metasurfaces with maximum-chiral transmission resonances stable against incident beam tilting. We have explained the origin of substrate-induced optical chirality in terms of the resonant state expansion theory and verified it for several illustrative examples. We believe our approach can be implemented for various planar structures and different material platforms, as well as it can be applied to stacks of planar chiral metasurfaces made of different transparent materials and separately interacting with waves of certain helicities and wavelengths.  

\medskip

\section*{Supporting Information}

Supporting Information is available from the authors upon reasonable request.

\medskip


\section*{Acknowledgements} \par 
The authors are grateful to Ivan Toftul and Qinghai Song for many useful discussions. The work of M.V.G., A.A.A and A.V.M was supported by the Russian Science Foundation (project 23-42-00091, https://rscf.ru/project/23-42-00091/).  Y.K. acknowledges a support of the Australian Research Council (Grant No. DP210101292) and the International Technology Center Indo-Pacific (ITC IPAC) via Army Research Office (contract FA520923C0023)

\section*{Author contributions} 

M.V.G. and A.A.A. conceived the original idea. A.A.A. and A.V.M. designed the PCS metasurfaces and performed numerical simulations. M.V.G and E.A.M developed an analytical RSE theory. M.V.G., A.A.A and Y.S.K. analyzed the data and interpreted the results. All authors contributed to the writing the manuscript.

\section*{Conflict of Interest} 

The authors declare no conflict of interests. 

\section*{Data Availability Statement}

The data that support the findings of this study are available from the
corresponding authors upon reasonable request.

%
%


\newpage

\begin{thebibliography}{10}
	\providecommand{\url}[1]{\texttt{#1}}
	\providecommand{\urlprefix}{URL }
	
	\bibitem{Herschel1822}
	J.~W.~F. Herschel,
	\newblock \emph{Transactions of the Cambridge Philosophical Society}
	\textbf{1822}, \emph{1} 43–50.
	
	\bibitem{Fresnel1824}
	A.-J. Fresnel,
	\newblock \emph{Bull. Sci. Soc. Philomath} \textbf{1824}, 147--158.
	
	\bibitem{Kelvin1894}
	W.~T.~B. Kelvin,
	\newblock \emph{The Molecular Tactics of a Crystal},
	\newblock Robert Boyle lecture. Clarendon Press, \textbf{1894}.
	
	\bibitem{Polavarapu2018}
	P.~Polavarapu,
	\newblock \emph{Chiral Analysis: Advances in Spectroscopy, Chromatography and
		Emerging Methods},
	\newblock Elsevier, Amsterdam, Netherlands, \textbf{2018}.
	
	\bibitem{AvalosOvando2022}
	O.~Ávalos Ovando, E.~Y. Santiago, A.~Movsesyan, X.-T. Kong, P.~Yu, L.~V.
	Besteiro, L.~K. Khorashad, H.~Okamoto, J.~M. Slocik, M.~A. Correa-Duarte,
	M.~Comesaña-Hermo, T.~Liedl, Z.~Wang, G.~Markovich, S.~Burger, A.~O.
	Govorov,
	\newblock \emph{ACS Photonics} \textbf{2022}, \emph{9}, 7 2219.
	
	\bibitem{He2018}
	C.~He, G.~Yang, Y.~Kuai, S.~Shan, L.~Yang, J.~Hu, D.~Zhang, Q.~Zhang, G.~Zou,
	\newblock \emph{Nature Communications} \textbf{2018}, \emph{9}, 1.
	
	\bibitem{Lodahl2017}
	P.~Lodahl, S.~Mahmoodian, S.~Stobbe, A.~Rauschenbeutel, P.~Schneeweiss,
	J.~Volz, H.~Pichler, P.~Zoller,
	\newblock \emph{Nature} \textbf{2017}, \emph{541}, 7638 473.
	
	\bibitem{Gansel2009}
	J.~K. Gansel, M.~Thiel, M.~S. Rill, M.~Decker, K.~Bade, V.~Saile, G.~von
	Freymann, S.~Linden, M.~Wegener,
	\newblock \emph{Science} \textbf{2009}, \emph{325}, 5947 1513.
	
	\bibitem{Singh2013}
	J.~H. Singh, G.~Nair, A.~Ghosh, A.~Ghosh,
	\newblock \emph{Nanoscale} \textbf{2013}, \emph{5}, 16 7224.
	
	\bibitem{Gibbs2013}
	J.~G. Gibbs, A.~G. Mark, S.~Eslami, P.~Fischer,
	\newblock \emph{Applied Physics Letters} \textbf{2013}, \emph{103}, 21 213101.
	
	\bibitem{Kaschke2014}
	J.~Kaschke, M.~Blome, S.~Burger, M.~Wegener,
	\newblock \emph{Optics Express} \textbf{2014}, \emph{22}, 17 19936.
	
	\bibitem{Esposito2016}
	M.~Esposito, V.~Tasco, F.~Todisco, M.~Cuscun{\`{a}}, A.~Benedetti, M.~Scuderi,
	G.~Nicotra, A.~Passaseo,
	\newblock \emph{Nano Letters} \textbf{2016}, \emph{16}, 9 5823.
	
	\bibitem{Plum2007}
	E.~Plum, V.~A. Fedotov, A.~S. Schwanecke, N.~I. Zheludev, Y.~Chen,
	\newblock \emph{Applied Physics Letters} \textbf{2007}, \emph{90}, 22 223113.
	
	\bibitem{Decker2010}
	M.~Decker, R.~Zhao, C.~M. Soukoulis, S.~Linden, M.~Wegener,
	\newblock \emph{Optics Letters} \textbf{2010}, \emph{35}, 10 1593.
	
	\bibitem{Dietrich2014}
	K.~Dietrich, C.~Menzel, D.~Lehr, O.~Puffky, U.~Hübner, T.~Pertsch,
	A.~Tünnermann, E.-B. Kley,
	\newblock \emph{Applied Physics Letters} \textbf{2014}, \emph{104}, 19 193107.
	
	\bibitem{Gorkunov2014}
	M.~V. Gorkunov, A.~A. Ezhov, V.~V. Artemov, O.~Y. Rogov, S.~G. Yudin,
	\newblock \emph{Applied Physics Letters} \textbf{2014}, \emph{104}, 22 221102.
	
	\bibitem{Kondratov2016}
	A.~V. Kondratov, M.~V. Gorkunov, A.~N. Darinskii, R.~V. Gainutdinov, O.~Y.
	Rogov, A.~A. Ezhov, V.~V. Artemov,
	\newblock \emph{Physical Review B} \textbf{2016}, \emph{93}, 19 195418.
	
	\bibitem{Zhu2017}
	A.~Y. Zhu, W.~T. Chen, A.~Zaidi, Y.-W. Huang, M.~Khorasaninejad, V.~Sanjeev,
	C.-W. Qiu, F.~Capasso,
	\newblock \emph{Light: Science {\&} Applications} \textbf{2017}, \emph{7}, 2
	17158.
	
	\bibitem{Gorkunov2018}
	M.~V. Gorkunov, O.~Y. Rogov, A.~V. Kondratov, V.~V. Artemov, R.~V. Gainutdinov,
	A.~A. Ezhov,
	\newblock \emph{Scientific Reports} \textbf{2018}, \emph{8}, 1 11623.
	
	\bibitem{Tanaka2020}
	K.~Tanaka, D.~Arslan, S.~Fasold, M.~Steinert, J.~Sautter, M.~Falkner,
	T.~Pertsch, M.~Decker, I.~Staude,
	\newblock \emph{{ACS} Nano} \textbf{2020}, \emph{14}, 11 15926.
	
	\bibitem{Gorkunov2020}
	M.~V. Gorkunov, A.~A. Antonov, Y.~S. Kivshar,
	\newblock \emph{Physical Review Letters} \textbf{2020}, \emph{125}, 9 093903.
	
	\bibitem{Overvig2021}
	A.~Overvig, N.~Yu, A.~Al{\`{u}},
	\newblock \emph{Physical Review Letters} \textbf{2021}, \emph{126}, 7 073001.
	
	\bibitem{Gorkunov2021}
	M.~V. Gorkunov, A.~A. Antonov, V.~R. Tuz, A.~S. Kupriianov, Y.~S. Kivshar,
	\newblock \emph{Advanced Optical Materials} \textbf{2021}, \emph{9} 2100797.
	
	\bibitem{Zhang2022}
	X.~Zhang, Y.~Liu, J.~Han, Y.~Kivshar, Q.~Song,
	\newblock \emph{Science} \textbf{2022}, \emph{377}, 6611 1215.
	
	\bibitem{Chen2023}
	Y.~Chen, H.~Deng, X.~Sha, W.~Chen, R.~Wang, Y.-H. Chen, D.~Wu, J.~Chu, Y.~S.
	Kivshar, S.~Xiao, C.-W. Qiu,
	\newblock \emph{Nature} \textbf{2023}, \emph{613}, 7944 474.
	
	\bibitem{Kuehner2023}
	L.~Kühner, F.~J. Wendisch, A.~A. Antonov, J.~Bürger, L.~Hüttenhofer,
	L.~de~S.~Menezes, S.~A. Maier, M.~V. Gorkunov, Y.~Kivshar, A.~Tittl,
	\newblock \emph{Light: Science \& Applications} \textbf{2023}, \emph{12}, 1.
	
	\bibitem{Fernandez-Corbaton2016}
	I.~Fernandez-Corbaton, M.~Fruhnert, C.~Rockstuhl,
	\newblock \emph{Physical Review X} \textbf{2016}, \emph{6}, 3 031013.
	
	\bibitem{Powell2010}
	D.~A. Powell, Y.~S. Kivshar,
	\newblock \emph{Applied Physics Letters} \textbf{2010}, \emph{97}, 9.
	
	\bibitem{Albooyeh2015}
	M.~Albooyeh, R.~Alaee, C.~Rockstuhl, C.~Simovski,
	\newblock \emph{Physical Review B} \textbf{2015}, \emph{91}, 19 195304.
	
	\bibitem{Tian2020}
	J.~Tian, Q.~Li, P.~A. Belov, R.~K. Sinha, W.~Qian, M.~Qiu,
	\newblock \emph{ACS Photonics} \textbf{2020}, \emph{7}, 6 1436.
 
        \bibitem{Toftul2024}
        I.~Toftul, P.~Tonkaev, K.~Koshelev, F.~Lai, Q.~Song, M.~Gorkunov, Y.~Kivshar
        \newblock \emph{arXiv:2406.11300} \textbf{2024}.

 	\bibitem{Semnani2020}
	B.~Semnani, J.~Flannery, R.~Al~Maruf, M.~Bajcsy,
	\newblock \emph{Light: Science \& Applications} \textbf{2020}, \emph{9}, 1 23.
	
	\bibitem{Shi2022}
	T.~Shi, Z.-L. Deng, G.~Geng, X.~Zeng, Y.~Zeng, G.~Hu, A.~Overvig, J.~Li, C.-W.
	Qiu, A.~Alù, Y.~S. Kivshar, X.~Li,
	\newblock \emph{Nature Communications} \textbf{2022}, \emph{13}, 1.
	
	\bibitem{Gorkunov2024}
	M.~V. Gorkunov, A.~A. Antonov,
	\newblock \emph{Rational design of maximum chiral dielectric metasurfaces},
	243--286,
	\newblock Elsevier,
	\newblock ISBN 9780323951951, \textbf{2024}.

 	\bibitem{Muljarov2010}
	E.~A. Muljarov, W.~Langbein, R.~Zimmermann,
	\newblock \emph{EPL (Europhysics Letters)} \textbf{2010}, \emph{92}, 5 50010.

 	\bibitem{Muljarov2018}
	E.~A. Muljarov, T.~Weiss, 
	\newblock \emph{Optics Letters} \textbf{2018}, \emph{43}, 9 1978.

  	\bibitem{Weiss2017}
	T.~Weiss, M.~Schäferling, H.~Giessen, N.~A.~Gippius, S.~G.~Tikhodeev, W.~Langbein, E.~A.~Muljarov,
	\newblock \emph{Physical Review B} \textbf{2017}, \emph{96}, 4 045129.
 
	\bibitem{Both2022}
	S.~Both, M.~Schäferling, F.~Sterl, E.~A. Muljarov, H.~Giessen, T.~Weiss,
	\newblock \emph{ACS Nano} \textbf{2022}, \emph{16}, 2 2822.

 	\bibitem{Almousa2023}
	S.~F. Almousa, E.~A. Muljarov,
	\newblock \emph{Physical Review B} \textbf{2023}, \emph{107}, 8 L081401.

 	\bibitem{Almousa2024}
	S.~F.~Almousa, T.~Weiss, E.~A.~Muljarov,
	\newblock \emph{Physical Review B} \textbf{2024}, \emph{109}, 4 L041410.

 	\bibitem{Sakoda1995}
	K.~Sakoda,
	\newblock \emph{Physical Review B} \textbf{1995}, \emph{52}, 11 7982.
	
	\bibitem{Hopkins2015}
	B.~Hopkins, A.~N. Poddubny, A.~E. Miroshnichenko, Y.~S. Kivshar,
	\newblock \emph{Laser {\&} Photonics Reviews} \textbf{2015}, \emph{10}, 1 137.
	
	\bibitem{Doiron2022}
	C.~F. Doiron, I.~Brener, A.~Cerjan,
	\newblock \emph{Nature Communications} \textbf{2022}, \emph{13}, 1.
	
	\bibitem{Zhen2014}
	B.~Zhen, C.~W. Hsu, L.~Lu, A.~D. Stone, M.~Soljačić,
	\newblock \emph{Physical Review Letters} \textbf{2014}, \emph{113}, 25 257401.
	
	\bibitem{Hsu2017}
	C.~W. Hsu, B.~Zhen, M.~Soljačić, A.~D. Stone 
 \newblock \emph{arXiv:1708.02197} \textbf{2017}.

 	\bibitem{Weiss2018}
	T.~Weiss, E.~A. Muljarov,
	\newblock \emph{Physical Review B} \textbf{2018}, \emph{98}, 8 085433.

	\bibitem{Sztranyovszky2023}
	Z.~Sztranyovszky, W.~Langbein, E.~A. Muljarov,
	\newblock \emph{Physical Review Research} \textbf{2023}, \emph{5}, 1 013209.
	
	\bibitem{Fan2003}
	S.~Fan, W.~Suh, J.~D. Joannopoulos,
	\newblock \emph{Journal of the Optical Society of America A} \textbf{2003},
	\emph{20}, 3 569.
	
	\bibitem{Kim2021a}
	K.~Kim, J.~Kim,
	\newblock \emph{Advanced Optical Materials} \textbf{2021}, \emph{9}, 22.
	
	\bibitem{Zong2022}
	S.~Zong, D.~Zeng, G.~Liu, Y.~Wang, Z.~Liu, J.~Chen,
	\newblock \emph{Optics Express} \textbf{2022}, \emph{30}, 22 40470.
	
	\bibitem{Liu2023}
	Q.-K. Liu, Y.~Li, Z.~Lu, Y.~Zhou, W.~M. Liu, X.-Q. Luo, X.-L. Wang,
	\newblock \emph{Physical Review B} \textbf{2023}, \emph{108}, 15 155410.
	
	\bibitem{Luo2023}
	X.~Luo, X.~Du, R.~Huang, G.~Li,
	\newblock \emph{Laser {\&} Photonics Reviews} \textbf{2023}, \emph{17}, 10.
	
	\bibitem{Gryb2023}
	D.~Gryb, F.~J. Wendisch, A.~Aigner, T.~Gölz, A.~Tittl, L.~de~S.~Menezes, S.~A.
	Maier,
	\newblock \emph{Nano Letters} \textbf{2023}, \emph{23}, 19 8891.
	
	\bibitem{Koshelev2023}
	K.~Koshelev, Y.~Tang, Z.~Hu, I.~I. Kravchenko, G.~Li, Y.~Kivshar,
	\newblock \emph{{ACS} Photonics} \textbf{2023}, \emph{10}, 1 298.
	
	\bibitem{Voronin2022}
	K.~Voronin, A.~S. Taradin, M.~V. Gorkunov, D.~G. Baranov,
	\newblock \emph{{ACS} Photonics} \textbf{2022}, \emph{9} 2652–2659.
	
	\bibitem{Decker2015}
	M.~Decker, I.~Staude, M.~Falkner, J.~Dominguez, D.~N. Neshev, I.~Brener,
	T.~Pertsch, Y.~S. Kivshar,
	\newblock \emph{Advanced Optical Materials} \textbf{2015}, \emph{3}, 6 813.
	
	\bibitem{Duan2023}
	Q.~Duan, Y.~Zeng, Y.~Yin, J.~Xu, Z.~Chen, Z.~Hao, H.~Chen, Y.~Liu,
	\newblock \emph{Photonics Research} \textbf{2023}, \emph{11}, 11 1919.
	
	\bibitem{Chen2021}
        Y.~Chen, W.~Chen, X.~Kong, D.~Wu, J.~Chu, C.-W.~Qiu,  Phys. Rev. Lett. 128, 146102 (2022).
        \newblock \emph{Physical Review Letters} \textbf{2022}, \emph{128}, 14 146102.
	
	\bibitem{Su2024}
	Z.~Su, Y.~Yang, B.~Xiong, R.~Zhao, Y.~Wang, L.~Huang,
	\newblock \emph{Advanced Optical Materials} \textbf{2024}, \emph{12}, 16 2303195.

	\bibitem{Koshelev2018}
	K.~Koshelev, S.~Lepeshov, M.~Liu, A.~Bogdanov, Y.~Kivshar,
	\newblock \emph{Physical Review Letters} \textbf{2018}, \emph{121}, 19 193903.
	
\end{thebibliography}
\end{document}